# Probing topological phase transition in silicene via Klein tunneling in a normal/ferromagnetic/normal silicene junction

Partha Goswami

Deshbandhu College, University of Delhi, Kalkaji, New Delhi-110019, India Tel:+91-9810163784

physicsgoswami@gmail.com

**Abstract**  The two apparently disparate phenomena, viz. the Klein tunneling and the electric field driven topological phase transition(TPT), exhibited by the silicene converge on the issue of the *no change in the pseudo-spin*. An ensuing possibility is that the former, in the case of a normal-magnetic–normal silicene junction, may be considered theoretically as a probe to ascertain the onset of the latter. In this communication, we explore this option and calculate the total transmission probability(TTP). We find that TPT is characterized by a cusp in a plot of TTP as a function of the electric field.

## 1. Introduction

The purpose of this short communication is to report the theoretical investigation of the Klein tunneling of Dirac fermions in a normal-magnetic-normal silicene junction [1] deployed for probing the out-of-plane electric field induced phase transition in silicene from a topological insulator to a valley-spin-polarized metal characterized by the by the spin-valley locking . While the magnetic silicene, characterized by the non-zero exchange field 'M', involves the presence of the applied electric field $E_z$ alluded to perpendicular to its plane, the normal silicene sheets have M = 0 and $E_z$ = 0. The exchange field arises due to proximity coupling to a ferro-magnet such as depositing Fe atoms to the silicene surface or depositing silicene to a ferromagnetic insulating substrate. The silicene is basically a semimetal because the valence and conduction bands touch at the Fermi level. It consists of a honeycomb lattice of silicon atoms with two sub-lattices made of A and B sites [2,3]. The states near the Fermi energy are π orbitals residing near the Dirac points **K** and **K′** [2,3]at opposite corners of the hexagonal Brillouin zone. The silicene sheet, in fact, has linear band crossing at the **K** and **K′** symmetry points. As in graphene, thus, the charge carriers in silicene behave like relativistic particles with a conical energy spectrum and Fermi velocity $v_F \approx 10^6$ m-s$^{-1}$. The honeycomb lattice of the system is distorted due to a larger (than carbon) ionic radius of silicon atom and forms a buckled structure. The A and B sites per unit cell form two sub-lattices separated by a perpendicular distance, say, 2ℓ. The structure generates a staggered sub-lattice potential 2 ℓ$E_z$ between silicon atoms at A sites and B sites for an applied electric field $E_z$.Silicene has a larger spin-orbit coupling (SOC)induced gap than graphene due to its buckled structure[4,5,6]. Remarkably, the silicene exhibits a tunable band gap due to the applied electric field $E_z$ perpendicular to the system-plane. Tuning of $E_z$, allows for rich behavior varying from a topological insulator(TI) to a band insulator(BI) with a valley spin-polarized metal (VSPM) at a critical value ($E_c$) in between. In fact, at the critical point with ($E_z/E_c$) = 1.00, the gap of one of the spin-split bands closes to give a Dirac point while at the other K point it is gapped [4,5,6]. Furthermore, it is the other spin-split band which has no gap. This is an out-of-plane electric field driven TI to theVSPM transition. Here-in-after we shall refer to this transition as the topological phase transition(TPT). A detection method of TPT by employing both the Friedel oscillation and collective excitation in silicene had been reported by Tabert et al.[7] and Chang et al[8]. The main thrust area of these authors was to calculate the dynamical polarization function and the undamped plasmon mode emerging from the single-particle excitation spectrum. As already stated, our aim is to present here an alternative (theoretical) tool, viz. the transmission probability in the scattering of Dirac electrons( Klein tunneling) in a normal-magnetic–normal silicene junction to ascertain the occurrence of TPT. We shall explain in section2, the convergence of two disparate phenomena,viz. the Klein tunneling and TPT, on the issue of the pseudo-spin preservation has basically propelled us to consider the idea of calculating the total transmission probability(TTP) at TPT. Our effort has been paid off as TPT is found to be characterized by a cusp in a plot of TTP as a function of the electric field. The communication ends with brief description of some additional facts relating to TPT.

## 2.Detection of TPT based on Klein tunneling

It must be mentioned here that the existence of Dirac fermions itself in silicene- an allotrope of silicon is somewhat speculative due to the lack of direct evidences from experiment [9,10]. Fleurence et al.[11] possibly have succeeded for the first time in preparing silicene on Si wafers via a conductive ceramic zirconium diboride ($ZrB_2$) buffer layer, though there have been quite a few independent silicene synthesis reports around the same time on the conducting substrate, such as that of the epitaxial silicene sheets on silver (111) [2,3,12]. We, however, assume that the physical reality of materials fabrication technique allows us to obtain 2D layers of this allotrope of silicon where the existence of Dirac fermions is undisputed reality. Accordingly, the Kane-Mele Hamiltonian [13] of the system, including both intrinsic and Rashba spin-orbit coupling, can be written in the following form. The dimensionless Hamitonian matrix around Dirac point $K_\xi$ ( the iso-spin index $\xi = \pm 1$) in the basis ($a_{\mathbf{k}\uparrow}$, $b_{\mathbf{k}\uparrow}$, $a_{\mathbf{k}\downarrow}$, $b_{\mathbf{k}\downarrow}$) in momentum space is h($\boldsymbol{\delta k}$)=[$\xi a\, \delta k_x\, (\gamma^5\, \gamma^0\, \gamma^x) + a\, \delta k_y\, (\gamma^5\, \gamma^0\, \gamma^y)$ ] + $\xi[t'_{soc}(\gamma^5\, \gamma^z\, \gamma^0\, \gamma^5) + \Delta_z\, (\gamma^5\, \gamma^z\, \gamma^0) + ai\, t'_{Rashba}\, (\gamma^z\, \delta k_x + \gamma^5\, \gamma^z \delta k_y)$]–$M\gamma^5$. Here 4X4 matrices($\gamma$) are in chiral basis. The first term is the kinetic energy. In a tight-binding approximation, the central term ( $-\Delta_\mathbf{z}\sum_{i,\sigma} \mu_i\ c^\dagger_{i\sigma}\, c_{i\sigma}$ ) with $\ell$ = 0.23Å, $\Delta_z = \ell E'_z$ ($E_z$ is the electric field) is the staggered sub-lattice potential term where $\mu_i = \pm 1$ for the A(B) site. These terms break the sub-lattice symmetry of the silicene's honey-comb structure and generate a gap. The exchange field **M** may arise due to coupling to a ferromagnet (FM) such as depositing Fe atoms to the silicene surface or depositing silicene to an FM insulating substrate. The terms $t'_{soc}(\gamma^5 \gamma^z \gamma^0 \gamma^5)$ and $ai\, t'_{Rashba}(\gamma^z \delta k_x + \gamma^5 \gamma^z \delta k_y)$ correspond to spin-orbit coupling. The Hamiltonian assumes much simpler form for the normal silicene ($\Delta_z = \ell E'_z = \frac{\ell E_z}{\left(\frac{\hbar v_F}{a}\right)} = 0$, and the exchange field M=0). Including the effect of the non-magnetic impurities [14], the Hamiltonian matrix in this case is given by

$$H^{(N)} = \begin{pmatrix} -\Delta_{soc}^N(a|\delta \mathbf{k}|) & \left(\frac{\hbar v_F}{a}\right) a \delta k_+ \\ \left(\frac{\hbar v_F}{a}\right) a \delta k_- & \Delta_{soc}^N(a\delta|\mathbf{k}|) \end{pmatrix}, \qquad (1)$$

where $\delta k_\pm = \xi\, \delta k_x \pm i\, \delta k_y$. The spin-split (the index $s_z = \pm 1$ for $\{\uparrow, \downarrow\}$) single-particle excitation spectrum is $\epsilon^{(N)}(a|\delta\mathbf{k}|) = E_{renorm} = \pm [(a|\delta\mathbf{k}|)^2 + \Delta_{soc}^{(N)}(a|\delta\mathbf{k}|)^2]^{1/2}$, the gap function is $\Delta_{soc}^{(N)}(a|\delta\mathbf{k}|) \equiv [\Delta_{soc}^2 - (1/16\, \acute{\Gamma}_\mathbf{k}^2)\{(a|\delta\mathbf{k}|)^2/(\Delta_{soc}^2 + (a|\delta\mathbf{k}|)^2)\}]^{1/2}$, $\Delta_{soc} = t'_{so} = \frac{t_{so}}{\left(\frac{\hbar v_F}{a}\right)}$, and $\xi = \pm 1$ around the valleys **K** and **K′**, respectively. The corresponding eigenvectors are given by

$$|\alpha_+\rangle = (1/\sqrt{2E_N E_{renorm}}) \begin{pmatrix} \left(\frac{\hbar v_F}{a}\right) a k_+ \\ E_N \end{pmatrix}, \quad |\alpha_-\rangle = (1/\sqrt{2E_N E_{renorm}}) \begin{pmatrix} \left(-\frac{\hbar v_F}{a}\right) a k_- \\ E_N \end{pmatrix}.$$

The wave function of the valley $\xi$ and the real spin $s_z$ for this region is $\psi_I(x<0) = \exp(ik_x x)\, |\alpha_+\rangle + r(\xi,s_z) \exp(-ik_x x)\, |\alpha_-\rangle$, where $r(\xi,s_z)$ is the reflection coefficient, and $E_N = E_{renorm} + \Delta_{soc}^{(N)}(a|\delta\mathbf{k}|)$. Since for the normal incidence, i.e. along x-axis in Figure 1 for which $\delta k_y = 0$, upon reflection, $k_x \to -k_x$ and $k_y \to k_y$, so $ak_+ \to -ak_-$. This is what we have done above. Upon replacing the Born approximation for scattering (by non-magnetic impurities) by the exact scattering cross-section for a single impurity, one obtains the reciprocal quasi-particle life-time $\acute{\Gamma}_\mathbf{k}^{-1}$. We notice from Eq.(1) that the velocity operator in the silicene case is given by $(V_x, V_y) = v_F(\xi \sigma_x, \sigma_y)$ where $\sigma = (\sigma_x, \sigma_y)$ is the pseudo-spin vector operator and $\xi = \pm 1$ around the valleys K and K′, respectively. As a consequence, for the normal incidence $\delta k_y = 0$ with the Hamiltonian in (4) we notice that the commutators $[H_{K(K')}, \xi \sigma_x] = 0$ in the absence of pseudo-spin-flip processes, and $[H_{K(K')}, \sigma_y] \neq 0$. So, indeed, $V_x$ is conserved conditionally (and $V_y$ is not conserved) around the valleys K and K′. Thus, the barrier remains always perfectly(imperfectly) transparent for angles close to the normal incidence in the absence(presence) of pseudo-spin-flip processes. This feature of the conservation of velocity component, unique to the mass-less Dirac fermions, is directly related to the perfect tunneling in Klein paradox.

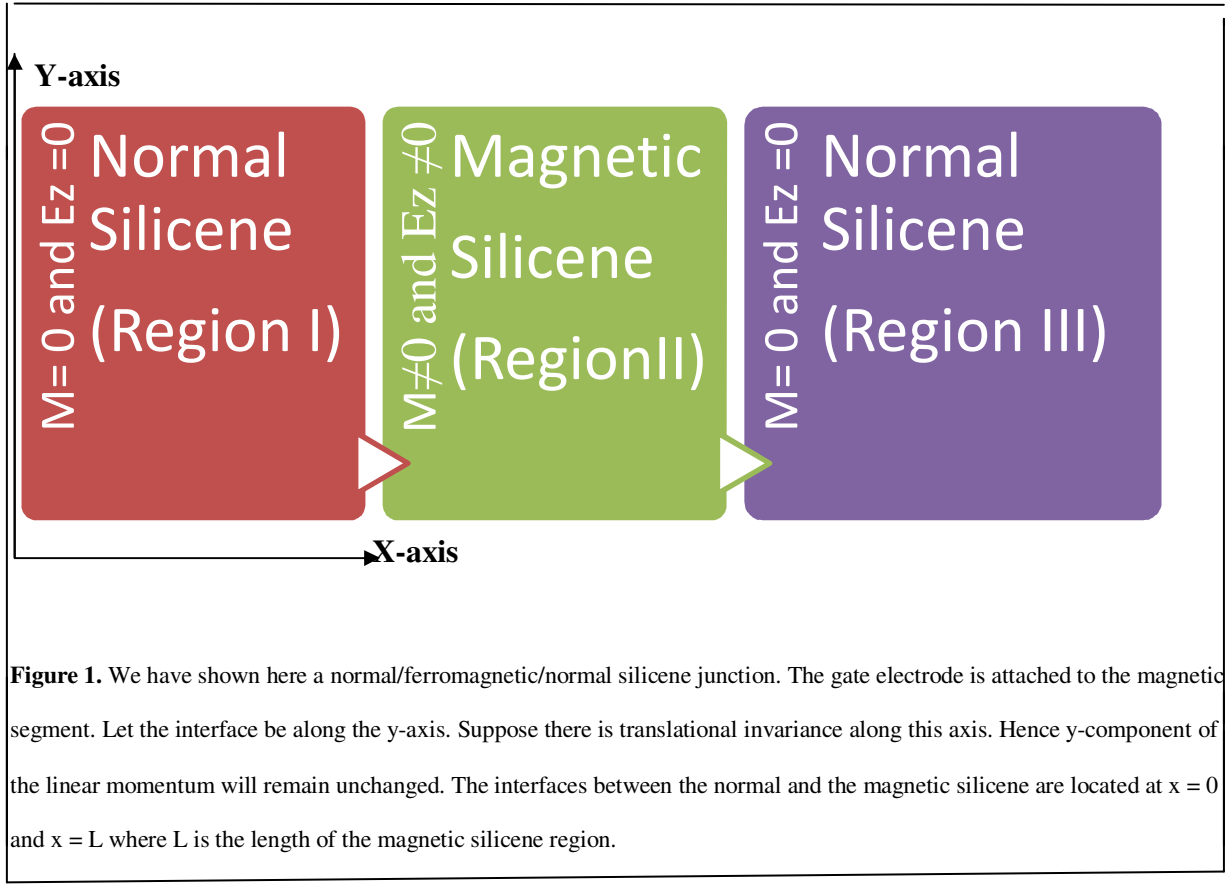

**Figure 1.** We have shown here a normal/ferromagnetic/normal silicene junction. The gate electrode is attached to the magnetic segment. Let the interface be along the y-axis. Suppose there is translational invariance along this axis. Hence y-component of the linear momentum will remain unchanged. The interfaces between the normal and the magnetic silicene are located at x = 0 and x = L where L is the length of the magnetic silicene region.

In the absence of the intrinsic Rashba terms $\sum_{\delta k} \{\xi a(\delta k_y + i\delta k_x) a^\dagger_{\delta k,\uparrow} a_{\delta k,\downarrow} + (\delta k_y - i\delta k_x) b^\dagger_{\delta k,\uparrow} b_{\delta k,\downarrow} + h.c.\}$ (we ignore this term as $t_2 \ll \Delta_{SOC}$) originating from the buckled honey-comb structure, we find the following eight bands from $h(\delta k)$: $\mathcal{E}(\delta k) = -s_z M \pm [(a|\delta k|)^2 + \{\xi s_z \Delta_{soc} + \Delta_z\}^2]^{1/2}$ where $\Delta_{soc} = t'_{so}$, $s_z = \pm 1$ for $\{\uparrow, \downarrow\}$. The effective staggered sub-lattice potential $V = \{\xi s_z \Delta_{soc} + \Delta_z\}$. The time reversal symmetry requires $\mathcal{E}(\xi, s_z, \delta k) = \mathcal{E}(\xi, s_z, -\delta k)$. The low-energy spectrum given above also comes up from the matrix $\hbar(\delta k,) = [\xi a \tau^0 \otimes \sigma^x \delta k_x + a\tau^0 \otimes \sigma^y \delta k_y + \xi \Delta_{soc} \tau^z \otimes \sigma^z + \Delta_z \tau^0 \otimes \sigma^z - M \sigma^0 \otimes \tau^z - (\mu) \tau^0 \otimes \sigma^0]$ where $\tau^i$ and $\sigma^i$, respectively, denote the Pauli matrices associated with the real spin and pseudo-spin of the Dirac electronic states. In view of this matrix, for the ferromagnetic silicene, it is possible to write down a phenomenological, minimal Hamiltonian matrix where only the pseudo-spin is in the foreground; the iso-spin(described by the index $\xi = \pm 1$) and the real spin (described by an index $s_z = \pm 1$) are in the background :

$$\hbar_{reduced}(\xi, s_z, \delta k) / \left(\frac{\hbar v_F}{a}\right) \approx \sum_{\delta k, sz}[(\xi s_z \Delta_{soc} + \Delta_z - s_z M) a^\dagger_{\delta k, sz} a_{\delta k, sz} + \{\xi a \delta k_x - ai \delta k_y\} a^\dagger_{\delta k, sz} b_{\delta k, sz} + (-\xi s_z \Delta_{soc} - \Delta_z - s_z M) b^\dagger_{\delta k, sz} b_{\delta k, sz} + \{\xi a \delta k_x + ai \delta k_y\} b^\dagger_{\delta k, sz} a_{\delta k, sz} - (\mu/\left(\frac{\hbar v_F}{a}\right))(a^\dagger_{\delta k, sz} a_{\delta k, sz} + b^\dagger_{\delta k, sz} b_{\delta k, sz})]. \quad (2)$$

Here $\mu' = (\mu/(\frac{\hbar v_F}{a}))$ is the dimensionless chemical potential of the fermion number. In the presence of the non-magnetic impurities, the eight bands $\mathcal{E}^{(M)}(a|\delta \mathbf{k}'|)$ for the ferromagnetic silicene are now given by

$$\mathcal{E}^{(M)}(a|\delta \mathbf{k}'|) = -s_z M \pm [\{(a|\delta \mathbf{k}'|)^2 + \Delta_{soc}^{(M)}(\xi, s_z, a|\delta \mathbf{k}'|)^2\}]^{1/2} - \mu', \qquad (3)$$

where $\mathcal{E}(\delta \mathbf{k}) = \mathcal{E}(-\delta \mathbf{k})$. The corresponding eigenvectors are now two-component Dirac spinor $|\gamma_{\pm}\rangle$:

$$|\gamma_+\rangle = \begin{pmatrix} \left(\frac{\hbar v_F}{a}\right) ak'_+ \\ E_M \end{pmatrix}, \quad |\gamma_-\rangle = \begin{pmatrix} \left(-\frac{\hbar v_F}{a}\right) ak'_- \\ E_M \end{pmatrix}, \qquad (4)$$

where

$$E_M = E_M(\xi, s_z, a|\delta \mathbf{k}'|) = E(a|\delta \mathbf{k}'|) + \Delta_{soc}^{(M)}(\xi, s_z, a|\delta \mathbf{k}'|) + s_z M + \mu',$$

$$E(a|\delta \mathbf{k}'|) = -s_z M \pm [\{(a|\mathbf{k}'|)^2 + \Delta_{soc}^{(M)}(\xi, s_z, a|\mathbf{k}'|)^2\}]^{1/2} - \mu', \qquad (5)$$

$$\Delta_{soc}^{(M)}(\xi, s_z, a|\delta \mathbf{k}'|) \equiv [(\xi s_z \Delta_{soc} + \Delta_z)^2 - (1/16\,\acute{\Gamma}_\mathbf{k}^2)\{(a|\delta \mathbf{k}'|)^2/((\xi s_z \Delta_{soc} + \Delta_z)^2 + (a|\delta \mathbf{k}'|)^2)\}]^{1/2}, \qquad (6)$$

$$(a\delta k'_x) = [(E(a|\delta \mathbf{k}|) + s_z M + \mu')^2 - \Delta_{soc}^{(M)}(\xi, s_z, a|\delta \mathbf{k}'|)^2 - (a\delta k_y)^2]^{1/2}, \quad \delta k'_\pm = \xi \delta k'_x \pm i\, \delta k_y. \qquad (7)$$

The wave function of the valley $\xi$ and the real spin $s_z$, corresponding to the region II ($0 \le x \le L$), now could be written as $\psi_{II}(0 \le x \le L, \xi, s_z) = A(\xi, s_z) \exp(ik_x' x) |\gamma_+\rangle + B(\xi, s_z) \exp(-ik_x' x) |\gamma_-\rangle$. Equations (5), (6) and (7) are the equation to determine $(ak'_x)$ iteratively. It is, thus, possible to describe electrons by the effective two-component wave function. It is also possible to calculate almost all the properties of silicene with this description. The wave function of the valley $\xi$ and the real spin $s_z$, corresponding to the region III ($x > L$), similarly could be written as $\psi_{III}(x > L, \xi, s_z) = t(\xi, s_z)(1/\sqrt{2E_N E_{renorm}}) \exp(ik_x x) |\alpha_+\rangle$ where $t(\xi, s_z)$ is the transmission coefficient. From the continuity of the wave function we obtain equations $\psi_I(x=0, y, \xi, s_z) = \psi_{II}(x=0, y, \xi, s_z)$, $\psi_{II}(x=L, y, \xi, s_z) = \psi_{III}(x=L, y, \xi, s_z)$. Unlike the conduction electrons in metals and semi-conductors described by the Schrödinger equation, we need to match the wave functions and not their derivatives. These equations enable us to determine the values of $A(\xi, s_z)$ and $B(\xi, s_z)$ occuring in $\psi_{II}(0 \le x \le L, \xi, s_z)$. The barrier transmission probability is given by $[t^*(\xi, s_z, a|\mathbf{k}|) t(\xi, s_z, a|\mathbf{k}|)] = 1 - \{r^*(\xi, s_z, a|\mathbf{k}|) r(\xi, s_z, a|\mathbf{k}|)\}$. Our aim is to calculate the total transmission probability below.

For the valley $\xi$ and the real spin $s_z$, the wave function continuity condition $\psi_I(x=0) = \psi_{II}(x=0)$ yields two equations $N(ak_+ - r(\xi, s_z) ak_-) = A(\xi, s_z)(ak'_+) - B(\xi, s_z)(ak'_-)$, and $N E_N (1 + r(\xi, s_z)) = (A(\xi, s_z) + B(\xi, s_z)) E_M$, where $N = (1/\sqrt{2E_N E_{renorm}})$. Similarly, the other continuity condition $\psi_{II}(x=L) = \psi_{III}(x=L)$ yields two equations $A(\xi, s_z) \exp(ik_x' L)(ak'_+) - B(\xi, s_z) \exp(-ik_x' L)(ak'_-) = N t(\xi, s_z) \exp(ik_x L)(ak_+)$, and $A(\xi, s_z) \exp(ik_x' L) E_M + B(\xi, s_z) \exp(-ik_x' L) E_M = N t(\xi, s_z) \exp(ik_x L) E_N$. From these continuity conditions we find

$$\frac{B(\xi,s_z)}{A(\xi,s_z)} = \frac{[(ak'_+)(1 + r(\xi,s_z))\varepsilon - (ak_+ - r(\xi,s_z)\,ak_-)]}{[(ak'_-)(1 + r(\xi,s_z))\varepsilon + (ak_+ - r(\xi,s_z)\,ak_-)]}, \tag{8}$$

We also find from these conditions

$$\frac{B(\xi,sz)}{A(\xi,sz)} = \frac{[\varepsilon\,(ak'_+) - (ak_+)]}{[\varepsilon\,(ak'_-) + (ak_+)]} \times \exp(-2i\,k_x'L). \tag{9}$$

We have put $(E_N/E_M\,(\xi,s_z, a|\mathbf{k}|)) = \varepsilon\,(\xi,s_z, a|\mathbf{k}|)$ above. Upon equating the right-hand-sides of (8) and (9) we obtain

$$\exp(ik_x'L)\frac{[(\varepsilon ak'_+ + ak_-)(1 + r(\xi,s_z)) - (ak_+ + ak_-)]}{[(\varepsilon ak'_- - ak_-)(1 + r(\xi,s_z)) + (ak_+ + ak_-)]} = \exp(-ik_x'L)\frac{(\varepsilon ak'_+ - ak_+)}{(\varepsilon ak'_- + ak_+)}. \tag{10}$$

Equation (10) eventually yields

$$(1 + r(\xi,s_z\,a|\mathbf{k}|))$$

$$= \frac{[\exp(ik_x'L)(ak_+ + ak_-)(\varepsilon(\xi,s_z,a|\mathbf{k}'|)\,ak'_+ + ak_+) + \exp(-ik_x'L)(ak_+ + ak_-)(\varepsilon(\xi,s_z,a|\mathbf{k}'|)\,ak'_+ - ak_+)]}{[\exp(ik_x'L)\,(\varepsilon(\xi,s_z,a|\mathbf{k}'|)\,ak'_+ + ak_-)(\varepsilon(\xi,s_z,a|\mathbf{k}'|)\,ak'_- + ak_+) - \exp(-ik_x'L)\,(\varepsilon\,ak'_+ - ak_+)(\varepsilon\,ak'_- - ak_-)]}. \tag{11}$$

The barrier transmission probability density is given by $[t^*(\xi,s_z, a|\delta\mathbf{k}|)\,t(\xi,s_z, a|\delta\mathbf{k}|)] = 1 - [r^*(\xi,s_z, a|\delta\mathbf{k}|)\,r(\xi,s_z, a|\delta\mathbf{k}|)]$. The total transmission probability (TTP) is obtained by the $\delta\mathbf{k}$-summation. For this purpose, we first divide the $\delta\mathbf{k}$- space into finite number of rectangular patches. We next determine the numerical values corresponding to each of these patches of the momentum-dependent density $[t^*(\xi,s_z, a|\delta\mathbf{k}|)\,t(\xi,s_z, a|\delta\mathbf{k}|)]$ and sum these values. We have generated these values through the surface plot above. These are illustrated in Fig. 2, where charge carriers from the "hot(red)" branch of the contour plot could be scattered into states within the same "hot(red)" branch but could not be transformed into any state on the "cold(blue)" branch. The matching between directions of pseudo-spin σ for quasi-particles inside and outside the barrier results in perfect tunneling. One can understand this perfect tunneling in terms of the conservation of pseudo-spin. Indeed, in the absence of pseudo-spin-flip processes, an electron moving to the right can be scattered only to a right-moving electron state or left-moving hole state.

Furthermore, we already have noted that in the VSPM transition case the dispersion relations are, for **K**, $E_{renorm}\,(\delta\mathbf{k}, s_z = -1, \xi = +1) \approx \pm(a|\delta\mathbf{k}|)$, and $E_{renorm}\,(\delta\mathbf{k}, s_z = +1, \xi = +1) \approx \pm[\{(a|\delta\mathbf{k}|)^2 + 4\Delta_z^2\}]^{\frac{1}{2}}$ for $\mu' = 0$. Similarly, for **K′**, $E_{renorm}\,(\delta\mathbf{k}, s_z = +1, \xi = -1) \approx \pm(a|\delta\mathbf{k}|)$, and $E_{renorm}\,(\delta\mathbf{k}, s_z = -1, \xi = -1) \approx \pm[\{(a|\delta\mathbf{k}|)^2 + 4\Delta_z^2\}]^{\frac{1}{2}}$ for $\mu' = 0$. Upon assuming that the valley states somehow correspond to real spins we find that states around **K** and **K′** are linked by a symmetry

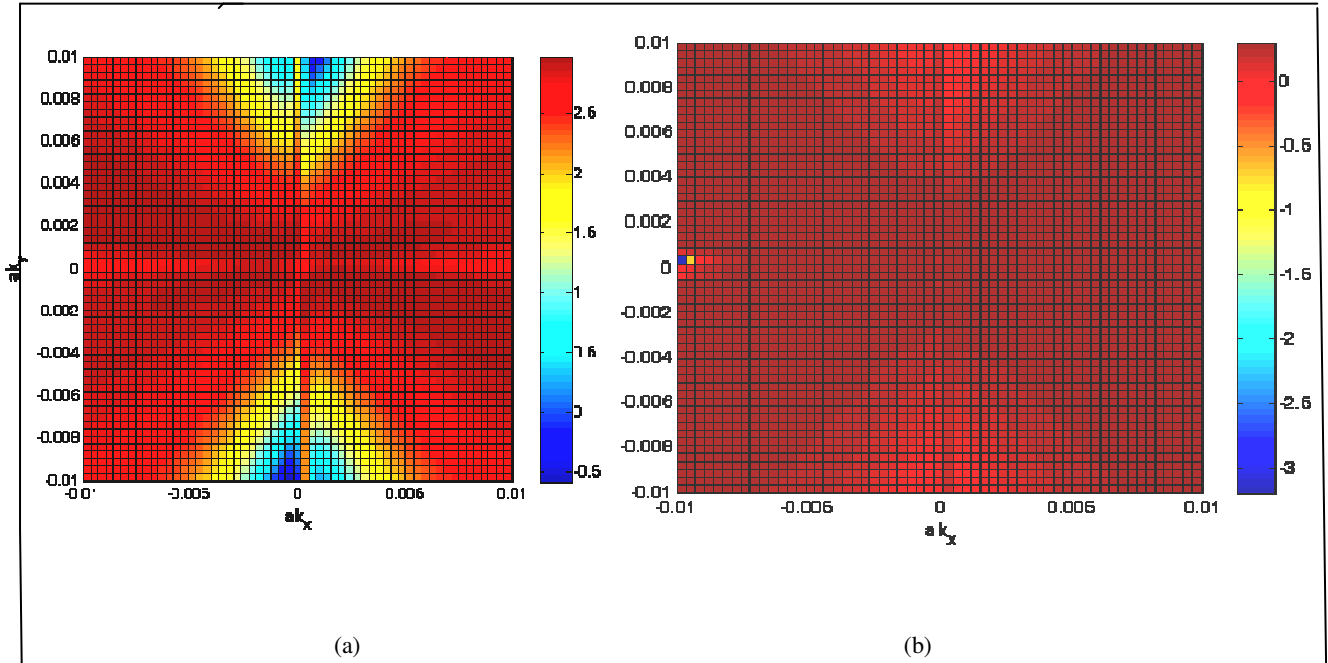

(a)　　　　　　　　　　　　　　(b)

**Figure 2:** In this figure we have demonstrated the outcome of the Klein-like quantum tunneling in the silicene system. The transmission probability (T) through a 20$a$-wide barrier is plotted as a function of the ($a\delta k_x$, $a\delta k_y$) in the case when the electric field is critical leading to VSPM. The figures correspond to the cases ($\xi = +1, s_z = -1, ph = +1$) and ($\xi = -1, s_z = +1, ph = +1$). The grading 'ph' is for particle-hole where for the particle ph = +1 and for the hole ph = −1. The figure (a) (the electric field ratio = 1.00) corresponds to the case when the non-magnetic impurity potential is absent and the figure (b) (the electric field ratio = 0.90) when the non-magnetic impurity potential squared $|V_0|^2 = +M/100$. The total transmission probability is greater, viz. 0.8531, in the former case, while it is 0.8313 in the latter case. We conclude from above that the elastic scattering by very weak non-magnetic impurities does not bring about a fundamental change in the transmission probability. However, it does bring down the occurrence of the VSPM criticality at lower applied field.

akin to the time-reversal symmetry of real spins. This is referred to as the iso-spin symmetry. The atomically sharp scattering centres are known[15,16] to lead to the broken iso-spin symmetry. We note that the conduction and the valence band operator formulation of the problem is more desirable and direct as the carriers in the system belong to these bands. The conduction and the valence bands, however, could be associated with the both the valleys **K** and **K′**. Now if there are mass-less Dirac fermions of certain real spin variety, say spin-up state, residing at **K**, after inter-valley scattering these particles transform into the massive Dirac particles of the same spin variety at **K′** rendering them 'differently pseudo-spinned'. The Dirac fermions of opposite spin variety at **K′** get transformed in a similar manner at **K** upon undergoing the scattering. This scenario is evidently non-convergent with that for Klein tunneling shown in Figure 2. Never-the-less, we are able to see from the expressions of $E_{renorm}$ ($\delta\mathbf{k}$, $s_z = -1$, $\xi = +1$),

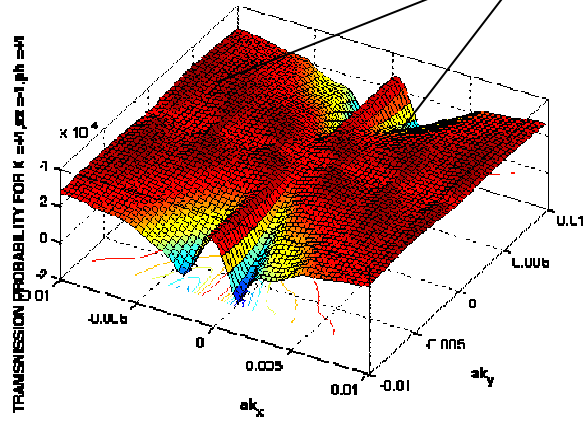

The patches correspond to resonance conditions

**Figure 3.** The 3D plot of the transmission probability (T) density through a D = 1000$a$-wide barrier as a function of the ($a\delta k_x$, $a\delta k_y$) in the case when the electric field is critical (leading to VSPM). Equations (11) yields that under resonance conditions $\delta k_x D = \pi N$, $N = 0, \pm 1, \ldots$ the barrier becomes transparent (T ≈ 1). The barrier remains always perfectly transparent for angles close to the normal incidence.

$E_{renorm}$ ($\delta\mathbf{k}$, $s_z$ = +1, $\xi$ = +1) etc., that close to the VSPM transition the real spin-flip process keeps the pseudo-spin unchanged when the inter-valley scattering takes place. The convergence of the two disparate phenomena, viz. the Klein tunneling and the VSPM transition, on the issue of the pseudo-spin invariance has caused us to move forward with the idea of the total transmission probability as a (theoretical) tool in the scattering of Dirac electrons in a normal-magnetic–normal silicene junction to ascertain the occurrence of TPT (or the VSPM transition). Our effort has been paid off as in a plot of the dependence of the total transmission probability in the electric field ratio without and with the inclusion of the effect of the non-magnetic impurities, respectively, the TPT (or, TI-VSPM-BI transition) is characterized by a precise cusp or smeared out bends (see Figure 4).

## 3. Additional facts relating to TPT

The Bloch fermions in silicene carry the orbital magnetic moment (M)[17] due to the self-rotation of the electrons as wave packets around the centre of mass, in addition to the spin. So, the magnetic moment in silicene has both orbital and spin character. Under symmetry operations, the orbital moment transforms exactly like the Berry curvature in silicene [18]. The silicene system has both time-reversal and inversion symmetry. Therefore, the Berry curvature and the orbital moment is in general nonzero (see appendix A). Interestingly, by actual calculation as in ref.[18], the latter is found to be proportional to the expression of the Berry curvature of the conduction band. Since the valley index ξ determines the sign of the orbital magnetic moment, the latter may also be termed as the valley magnetic moment (VMM). We find that the increase in $E_z$ beyond the critical value leads to the valley magnetic moment reversal. This is estimated to be two times greater than that of graphene[17]. Therefore, an applied magnetic field is expected to elicit greater response from silicene. Naturally, silicene is a better options to realize valley polarization than graphene. In Figure5, we have plotted the valley magnetic moment as a function of the dimensionless electric field close to the Dirac point. We find that, the critical point with $(E_z/E_c) = 1.00$ is characterized by the VMM (M) sign reversal. The VMM, in fact, vanish everywhere except at the Dirac points where they diverge. Our investigation, thus, demonstrates the ability of silicene to assume positive and negative values of orbital magnetic moments driven by the electric field tuning. This represents a novel approach of maneuvering Berry-phase effects for applications in micro-electronics associated with 2D Dirac materials.

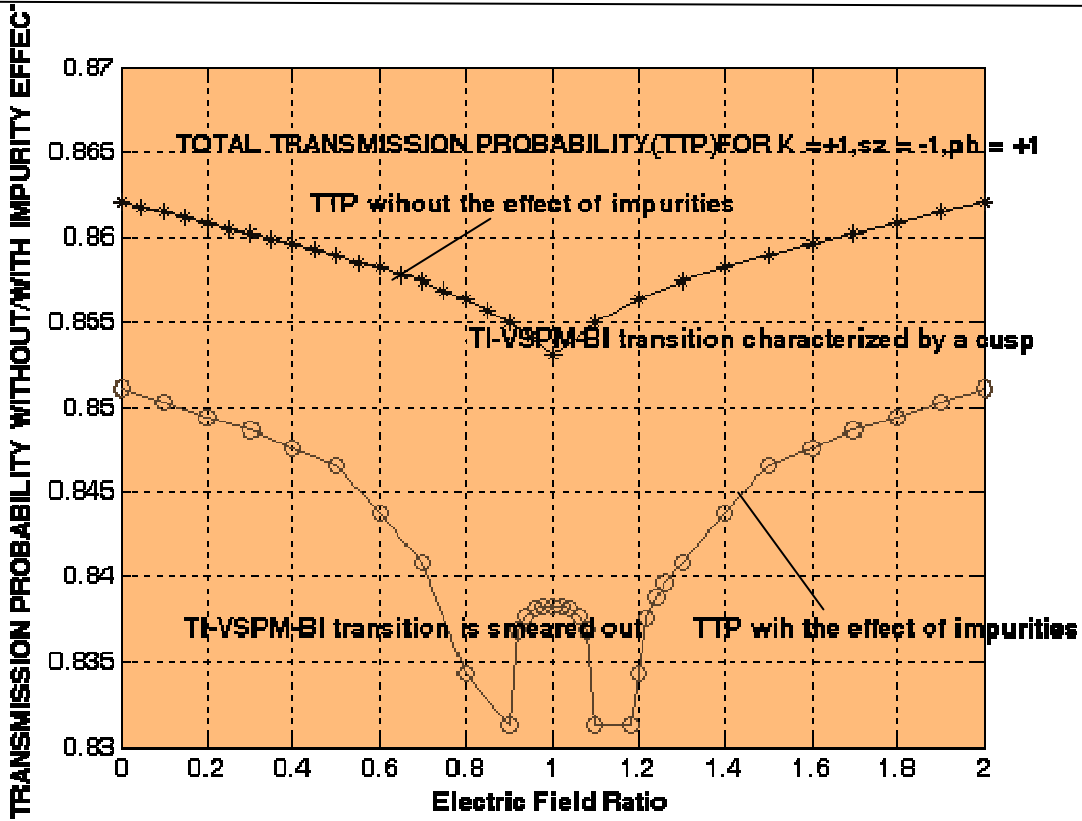

**Figure 4.** A plot of the total barrier transmission probability is given by [t*(ξ,s_z, a|**k**|) t(ξ,s_z, a|**k**|)] = 1− [ r*(ξ,s_z, a|**k**|) r(ξ,s_z, a|**k**|))] as a function of the electric field ratio. In the absence of the non-magnetic impurities, the TPT (or, TI-VSPM-BI transition) is characterized by a cusp, while in the presence of the impurities the tansition gets smeared out.

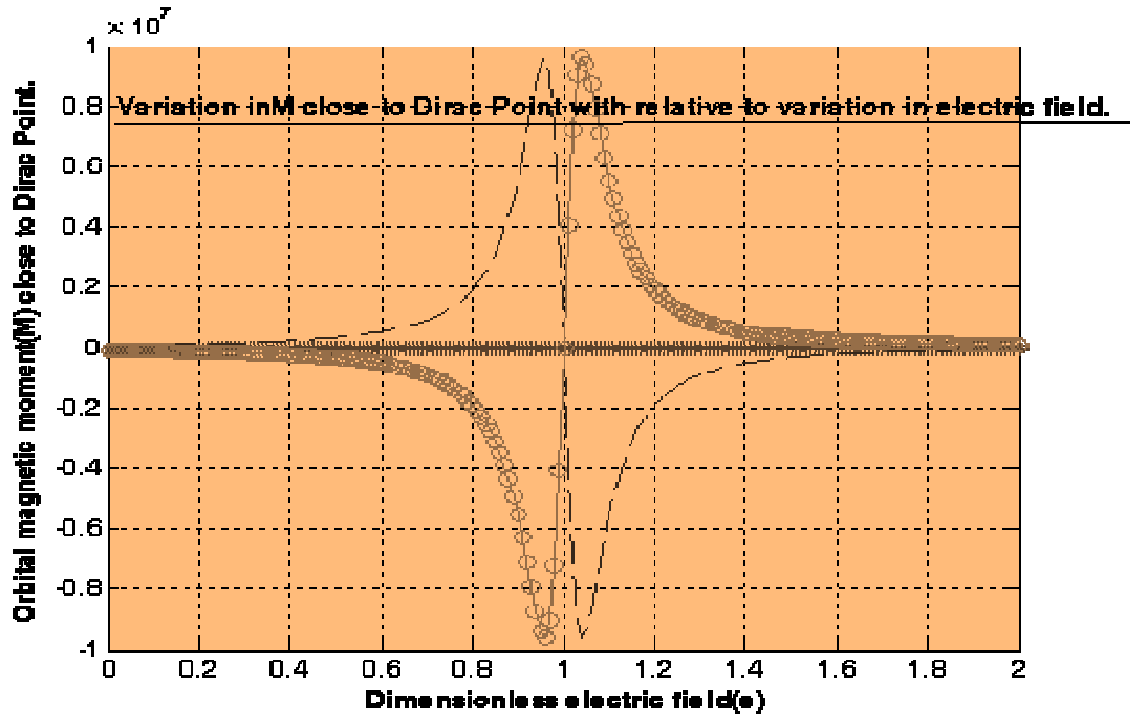

**Figure 5.** In this figure we have plotted the VMM (M) as a function of the dimensionless electric field(e) close to the Dirac point. We find that, the critical point with $(E_z/E_c) = 1.00$ is characterized by the VMM sign reversal.

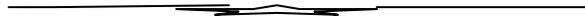

## Appendix A

In this appendix, we consider the wave packets in the semi-classical approach for spin and charge transportation of electrons in silicene. Our aim is to calculate the Berry curvature which has opposite signs in the two valleys **K** and **K′** for opposite spins as required by time-reversal (TR) symmetry. In the semi-classical approach electrons as wave packets are constructed from the Bloch waves. The physical origin of the Berry connection, and the curvature corresponds to the periodic part of the complex wave function $u_n(\mathbf{r}, \delta\mathbf{k})$ in the Bloch waves, where 'n' is the band index : For each band 'n' there is an associated connection given by $A_n(\delta\mathbf{k}) = i \int d^3\mathbf{r}\, u_n^*(\mathbf{r}, \delta\mathbf{k}) \nabla_\mathbf{k} u_n(\mathbf{r}, \delta\mathbf{k})$. The curvature, on the other hand, is given by $\Omega_n(\delta\mathbf{k}) = \nabla_\mathbf{k} \times A_n(\delta\mathbf{k})$. The Chern numbers(C) are defined as the integral of $\Omega_n(\delta\mathbf{k})$ over the whole Brillouin zone. Since, for our time-reversal invariant (TRI) silicene system ($\epsilon(\delta\mathbf{k}) = \epsilon(-\delta\mathbf{k})$), the wave functions $u_n(\mathbf{r}, \delta\mathbf{k})$ are defined up to an arbitrary phase, $A_n(\delta\mathbf{k})$ and $A_n(-\delta\mathbf{k})$ differ by a gauge transformation and $\Omega_n(\delta\mathbf{k}) = -\Omega_n(-\delta\mathbf{k})$. This implies that the Chern number $\iint_{BZ} \Omega_n(\delta\mathbf{k})\, d(\delta\mathbf{k})$ should be zero. We note that while time-reversal (T) is an unitary transformation, the space inversion(I) is anti-unitary and under the latter $\Omega_n(\delta\mathbf{k}) = \Omega_n(-\delta\mathbf{k})$. Since we have TI symmetry for the system under consideration, $\iint_R \Omega_n(\delta\mathbf{k})\, d(\delta\mathbf{k}) = n\pi$ for any region R. So, $\Omega_n$ must be either 0 or some delta functions, i.e. $\Omega_n(\delta\mathbf{k}) = \sum_i n_i\, \partial(\delta\mathbf{k} - \delta\mathbf{k}_i)\pi$. The $\partial$-functions are known as Berry fluxes. Since the Berry phase is only well defined up to mod $2\pi$, we shall have in general two types of Berry fluxes: 0 and π; a 'π' flux corresponds to a Dirac point here.

For the investigation at hand, the Berry curvature is determined by the eigenvalue problem $\bar{h}_{reduced}(\xi, s_z, \delta \mathbf{k}) / \left(\frac{\hbar v_F}{a}\right)$ $|u_n(\delta \mathbf{k})\rangle = \epsilon_{n,\mathbf{k}}|u_n(\delta \mathbf{k})\rangle$. We shall use below the notations $u_{K(K'),\pm}(\delta \mathbf{k})$ for the two-component eigenvectors representing the expansion coefficients of $|u_n(\delta \mathbf{k})\rangle$ around the two Dirac points $\mathbf{K}$ and $\mathbf{K'}$. We have seen in section 2 that the reduced, massive Dirac model hermitian Hamiltonian matrix for the silicene reads $\bar{h}_{reduced}(\xi, s_z, \delta \mathbf{k}) / \left(\frac{\hbar v_F}{a}\right) = [\xi a \sigma^x \delta k_x + a \sigma^y \delta k_y + V(\xi, s_z, a|\delta \mathbf{k}|)\sigma^z - (\mu / \left(\frac{\hbar v_F}{a}\right))\sigma^0]$, $V(\xi, s_z, a|\delta \mathbf{k}|) \approx \{\xi s_z \Delta_{soc} + \Delta_z\}$; $\mu' = (\mu/\left(\frac{\hbar v_F}{a}\right))$ is the dimensionless chemical potential of the fermion number. This gives the dispersion as $\epsilon(\delta \mathbf{k}) = \mu' + \frac{\epsilon(\delta \mathbf{k})_{\xi, s_z}}{\left(\frac{\hbar v_F}{a}\right)} \approx \pm\{(V(\xi, s_z, a|\delta \mathbf{k}|))^2 + (a|\delta \mathbf{k}|)^2\}^{1/2}$. The corresponding eigenvectors may be written as

$$u_{K,\pm}(\delta \mathbf{k}) = (1/\sqrt{2}) \begin{pmatrix} \sqrt{\left(\frac{\epsilon \pm V}{2\epsilon}\right)} \exp\left(-\frac{i\theta_k}{2}\right) \\ \pm\sqrt{\left(\frac{\epsilon \mp V}{2\epsilon}\right)} \exp\left(\frac{i\theta_k}{2}\right) \end{pmatrix}, \quad u_{K',\pm}(\delta \mathbf{k}) = (1/\sqrt{2}) \begin{pmatrix} \sqrt{\left(\frac{\epsilon \pm V}{2\epsilon}\right)} \exp\left(\frac{i\theta_k}{2}\right) \\ \mp\sqrt{\left(\frac{\epsilon \mp V}{2\epsilon}\right)} \exp\left(-\frac{i\theta_k}{2}\right) \end{pmatrix}.$$

Here $\theta_k = \arctan(k_y/k_x)$. The eigenvectors yield the connections, for example, $A_+(\mathbf{k}) = i u_+^\dagger(\mathbf{k}) \nabla_\mathbf{k} u_+(\mathbf{k})$. The corresponding curvature is $\Omega_+(\mathbf{k}) = -\text{Im}(N_{mr}/D_{mr})$ where $N_{mr} = [u_+^\dagger(\mathbf{k}) \nabla_\mathbf{k} \{\bar{h}_{reduced}(\xi, s_z, \delta \mathbf{k})/\left(\frac{\hbar v_F}{a}\right)\} u_-(\mathbf{k}) \times u_-^\dagger(\mathbf{k}) \nabla_\mathbf{k} \{\bar{h}_{reduced}(\xi, s_z, \delta \mathbf{k})/\left(\frac{\hbar v_F}{a}\right)\} u_+(\mathbf{k})]$, and $D_{mr} = (\epsilon_{+,\mathbf{k}} - \epsilon_{-,\mathbf{k}})^2$. We find the curvature for the conduction band in the form

$$\Omega_\xi(s_z, \Delta_z, a|\delta \mathbf{k}|) \sim (\xi/2) \times [V(\xi, s_z, a|\delta \mathbf{k}|)/\{(V(\xi, s_z, a|\delta \mathbf{k}|))^2 + (a|\delta \mathbf{k}|)^2\}^{3/2}],$$

$$V(\xi, s_z, a|\delta \mathbf{k}|) = \{\Delta_{soc}(a|\delta \mathbf{k}|) + \xi s_z \Delta_z\}, \quad \Delta_{soc}(a|\delta \mathbf{k}|) = (t'^2_{so} + (at'_2|\delta \mathbf{k}|)^2)^{1/2},$$

and $s_z = \pm 1$ for $\{\uparrow, \downarrow\}$. The Berry curvature has opposite signs in the two valleys $\mathbf{K}$ and $\mathbf{K'}$ for opposite spins as required by TR symmetry. It could be easily seen from above that for the 3D case if the dispersion is of the form $\epsilon(\mathbf{k}) = \pm\{(a\delta k_z)^2 + (a|\delta \mathbf{k}|)^2\}^{1/2}$ we have $\Omega_\pm(\mathbf{k}) = \mp(1/2) \mathbf{k}/k^3$ which means, if $\mathbf{k}$ is replaced by the real space vector $\mathbf{r}$, $\Omega_\pm(\mathbf{k})$ is in the form of a magnetic field due to the Dirac monopole with the magnetic charges ($\mp 1/2$). This is in agreement with the physical interpretation of Berry curvature as magnetic field in momentum space.